\documentclass{jpsj-suppl}
\usepackage{txfonts,bm} 

\newcommand{\Psfig}[2]{\includegraphics[width=#1]{#2}}

\newcommand{\SUN}[1]{\text{SU} ( #1 )}

\def\gev{\text{ GeV}}

\title{Compositeness of the $\Delta (1232)$ resonance in $\pi N$ scattering}

\author{Takayasu \textsc{Sekihara}$^{1}$, Takashi \textsc{Arai}$^{2}$, 
Junko \textsc{Yamagata-Sekihara}$^{3}$, Shigehiro \textsc{Yasui}$^{4}$}

\inst{$^{1}$Research Center for Nuclear Physics
  (RCNP), Osaka University, Ibaraki, Osaka, 567-0047, Japan \\
  $^{2}$Theory Center, IPNS, High Energy Accelerator Research
  Organization (KEK), 1-1 Oho, Tsukuba, Ibaraki 305-0801, Japan \\
  $^{3}$National Institute of Technology, Oshima College, Oshima,
  Yamaguchi, 742-2193, Japan \\
  $^{4}$Department of Physics, Tokyo Institute of Technology,
  Tokyo 152-8551, Japan}

\email{sekihara@rcnp.osaka-u.ac.jp}

\recdate{\today}

\abst{

  We evaluate the $\pi N$ compositeness of the $\Delta (1232)$
  resonance so as to clarify the internal structure of $\Delta (1232)$
  in terms of the $\pi N$ component.  Here the compositeness is
  defined as contributions from two-body wave functions to the
  normalization of the total wave function and is extracted from the
  $\pi N$ scattering amplitude.  In this study we employ the chiral
  unitary approach with the interaction up to the next-to-leading
  order plus a bare $\Delta$ term in chiral perturbation theory and
  describe $\Delta (1232)$ in an elastic $\pi N$ scattering.  Fitting
  the $\pi N$ scattering amplitude to the solution of the partial wave
  analysis, we obtain a large real part of the $\pi N$ compositeness
  for $\Delta (1232)$ comparable to unity and non-negligible imaginary
  part as well, with which we reconfirm the result in the previous
  study on the $\pi N$ compositeness for $\Delta (1232)$.

}

\kword{compositeness, chiral unitary approach, internal structure of
  $\Delta (1232)$}

\begin{document}
\maketitle

\section{Introduction}

The $\Delta (1232)$ resonance is one of the most fundamental hadrons
to understand the underlying theory of strong interaction, QCD.  The
most important influence on strong interaction is that $\Delta (1232)$
as a $| u \! \uparrow u \!  \uparrow u \! \uparrow \rangle$ state
leads to an idea that quarks have color degrees of
freedom~\cite{Han:1965pf}; otherwise, it breaks the Pauli principle
with respect to the exchange of quarks.  Moreover, $\Delta (1232)$ was
found to belong to a decuplet in the flavor $\SUN{3}$ symmetry
together with the $\Sigma (1385)$ and $\Xi (1530)$ resonances and it
predicted the existence and properties of the $\Omega ^{-}$ baryon,
which was followed by the experimental discovery.  These excellent
successes of the quark model for $\Delta (1232)$ and other decuplet
states strongly indicate that the decuplet states are described as
genuine $q q q$ states very well.

However, there are several suggestions that the effect of the
meson--nucleon cloud for $\Delta (1232)$ seems to be large.  For
instance, the $M1$ transition form factor for $\gamma ^{\ast} N \to
\Delta (1232)$ shows that the meson cloud effect brings $\sim 30 \%$
of the form factor at $Q^{2} = 0$~\cite{Sato:2009de}.  In addition,
the $\pi N$ component in $\Delta (1232)$ was studied in terms of the
so-called compositeness extracted from the $\pi N$ scattering
amplitude in a simple model~\cite{Aceti:2014ala}.  As a result, the
real part of the $\pi N$ compositeness is large and comparable to
unity although its imaginary part is non-negligible, which implies
large contribution of the $\pi N$ cloud to the internal structure of
$\Delta (1232)$.

In this study we aim at examining whether the $\pi N$ compositeness is
large or not in a more refined model for $\Delta (1232)$.  For this
purpose, we employ the so-called chiral unitary approach for the $\pi
N$ scattering~\cite{Kaiser:1995cy, Meissner:1999vr, Nieves:2001wt,
  Inoue:2001ip, Bruns:2010sv, Alarcon:2012kn, Garzon:2014ida}.  We
take the interaction kernel from chiral perturbation theory up to the
next-to-leading order plus a bare $\Delta$ term, and evaluate the loop
function in a dispersion relation.  We fit the model parameters to the
solution of the partial wave analysis for the $\pi N$ scattering
amplitude, and calculate the $\pi N$ compositeness for $\Delta (1232)$
from the $\pi N$ scattering amplitude.

\section{Framework}


\subsection{Compositeness from scattering amplitude}
\label{sec:2-1}

Recently the compositeness has been introduced into the hadron physics
so as to discuss the hadronic molecular component inside
hadrons~\cite{Hyodo:2011qc, Aceti:2012dd, Hyodo:2013nka,
  Sekihara:2014kya}.  The compositeness is defined as contributions
from two-body wave functions to the normalization of the total wave
function for the resonance, and corresponds to unity minus the field
renormalization constant intensively discussed in the
1960s~\cite{Weinberg:1962hj, Lurie:1963}.  Although the compositeness
is not observable and hence a model dependent quantity, it will be an
important piece of information on the structure of the resonance.

First we consider the scattering amplitude and compositeness in the
non-relativistic formulation, for simplicity.  The scattering
amplitude $T(E; \, \bm{q}^{\prime}, \, \bm{q})$, a solution of the
Lippmann--Schwinger equation, is described with the energy $E$ and
relative momenta in the initial and final states, $\bm{q}$ and
$\bm{q}^{\prime}$, respectively, and has a pole at $E = E_{\rm pole}$,
which coincides with the eigenenergy of the resonance state $| \Psi
\rangle$.  Near the resonance pole, the scattering amplitude is
dominated by the pole term in the expansion by the eigenstates of the
full Hamiltonian, and hence we have
\begin{equation}
T ( E ; \, \bm{q}^{\prime} , \, \bm{q} ) 
=
\langle \bm{q}^{\prime} | \hat{V} | \Psi \rangle
\frac{1}{E - E_{\rm pole}} 
\langle \Psi ^{\ast} | \hat{V} | \bm{q} \rangle ,
\end{equation}
where $\hat{V}$ is the operator of the interaction and $| \bm{q}
\rangle$ is the two-body state with relative momentum $\bm{q}$.  For
the bra vector of the resonance we take $\langle \Psi ^{\ast} |$
instead of $\langle \Psi |$, with which we can obtain the correct
normalization $\langle \Psi ^{\ast} | \Psi \rangle =
1$~\cite{Hyodo:2013nka, Sekihara:2014kya}.  Now we assume that the
interaction is separable type in general $L$-wave scattering
as done in Ref.~\cite{Aceti:2012dd}, which is essential to the correct
behavior of the amplitude near the threshold: $T_{L\text{-wave}} = |
\bm{q} |^{L} | \bm{q}^{\prime} |^{L} T^{\prime} ( E )$.  Then the
residue of the scattering amplitude becomes
$\langle \bm{q} | \hat{V} | \Psi \rangle
= 
\langle \Psi ^{\ast} | \hat{V} | \bm{q} \rangle
= g | \bm{q} |^{L}$
with the coupling constant of the resonance to the two-body state $g$.
As a result, the norm of the two-body wave function is calculated as
\begin{equation}
X \equiv \int \frac{d^{3} q}{( 2 \pi )^{3}} 
\langle \Psi ^{\ast} | \bm{q} \rangle
\langle \bm{q} | \Psi \rangle
= g^{2} \int \frac{d^{3} q}{( 2 \pi )^{3}} 
\frac{| \bm{q} |^{2 L}}
{\{ E_{\rm pole} - [M_{\rm th} + | \bm{q} |^{2} / (2 \mu)] \} ^{2}}
= - g^{2} \left [ \frac{d G_{L}}{d E} \right ]_{E = E_{\rm pole}} ,
\label{eq:X}
\end{equation}
where $M_{\rm th}$ and $\mu$ are the threshold of the two-body state
and the reduced mass, respectively, and we have used a relation
$\langle \bm{q} | \Psi \rangle = \langle \Psi ^{\ast} | \bm{q} \rangle
= g | \bm{q} |^{L} / \{ E_{\rm pole} - [ M_{\rm th} + | \bm{q} |^{2} /
(2 \mu ) ] \}$ obtained from $\langle \bm{q} | \hat{V} | \Psi \rangle
= \langle \Psi ^{\ast} | \hat{V} | \bm{q} \rangle = g | \bm{q} |^{L}$.
The $L$-wave loop function $G_{L} ( E )$ is defined as
\begin{equation}
G_{L} ( E ) \equiv \int \frac{d^{3} q}{( 2 \pi )^{3}} 
\frac{| \bm{q} |^{2 L}}{E - [ M_{\rm th} + | \bm{q} |^{2} / (2 \mu) ]} .
\end{equation}

\subsection{$\Delta (1232)$ in chiral unitary approach}
\label{sec:2-2}

Next let us formulate the $\pi N$ scattering amplitude in the chiral
unitary approach.  In this study we solve the following scattering
equation in an algebraic form for the elastic $\pi N$ scattering:
\begin{equation}
{T^{\prime}}_{I L}^{\pm} ( w ) = {V^{\prime}}_{I L}^{\pm} ( w ) 
+ {V^{\prime}}_{I L}^{\pm} ( w ) G_{L} ( w ) {T^{\prime}}_{I L}^{\pm} ( w ) 
= \frac{1}{1 / {V^{\prime}}_{I L}^{\pm} ( w ) - G_{L} ( w ) } ,
\label{eq:BS}
\end{equation}
with the center-of-mass energy $w$, the interaction kernel
${V^{\prime}}_{I L}^{\pm}$ and full amplitude ${T^{\prime}}_{I
  L}^{\pm}$ in isospin $I$, $L$ wave, and total angular momentum $J =
L \pm 1/2$, and the $L$-wave loop function $G_{L}$.  The interaction
kernel is taken from chiral perturbation theory up to the
next-to-leading order, i.e., the Weinberg--Tomozawa term $V_{\rm WT}$,
the $s$- and $u$-channel nucleon [$N (940)$] exchange terms $V_{s +
  u}$, and the contact next-to-leading order term $V_{2}$, plus a bare
$\Delta$ term $V_{\Delta}$: $V = V_{\rm WT} + V_{s + u} + V_{2} +
V_{\Delta}$.  This is projected to the eigenstate $I$, $L$, and $J = L
\pm 1/2$ to be $V_{I L}^{\pm}$, and then the momentum prefactor $|
\bm{q} |^{2 L}$ is picked out as $V_{I L}^{\pm} = | \bm{q} |^{2 L}
{V^{\prime}}_{I L}^{\pm}$.  Now ${V^{\prime}}_{I L}^{\pm}$ is a
function only of the center-of-mass energy $w$ and we use it as the
interaction kernel in the scattering equation~\eqref{eq:BS}.  On the
other hand, the loop function is evaluated in a dispersion relation
with the relative momentum $| \bm{q} |^{2 L}$ inside the integral as
\begin{equation}
G_{L} ( w ) \equiv \int _{s_{\rm th}}^{\infty} \frac{d s^{\prime}}{2 \pi} 
\frac{\rho (s^{\prime}) q ( s^{\prime} )^{2 L}}{s - s^{\prime}} 
= i \int \frac{d^{4} q}{( 2 \pi )^{4}}
\frac{| \bm{q} |^{2 L}}{[ (P - q)^{2} - m_{\pi}^{2} ] (q^{2} - M_{N}^{2}) } ,
\quad
\rho (s) \equiv \frac{q(s)}{4 \pi w} ,
\label{eq:G}
\end{equation}
where $s = w^{2}$, $P^{\mu} = ( w , \, \bm{0})$, $m_{\pi}$ and $M_{N}$
are the pion and nucleon masses, respectively, $q(s)$ is the
center-of-mass momentum, and $s_{\rm th} \equiv ( m_{\pi} + M_{N}
)^{2}$.  We note that we need two subtraction constants for the
$p$-wave loop function.  In this study we fix one of them so that the
nucleon mass stays physical, for which we require $G_{L} ( w =
M_{N} ) = 0$.  From the $\pi N$ scattering amplitude, we can extract
the $\pi N$ compositeness $X_{\pi N}$ for $\Delta (1232)$ and $N
(940)$ with the formula~\eqref{eq:X}~\cite{Aceti:2012dd} with
replacing the loop function $G_{L}$ with that evaluated in the
dispersion relation~\eqref{eq:G}.

In this construction we have seven model parameters for the $\pi N$
scattering amplitude: four from the low-energy constants in the
next-to-leading order interaction, the bare $\Delta$ mass, the bare
$\pi N$-$\Delta$ coupling constant, and one subtraction constant
$\tilde{A}$ in $p$ wave, which enters as $G_{L = 1} ( w ) = ( s -
M_{N}^{2} ) \tilde{A} + (\text{finite part})$.  They are determined from
the fitting to the $\pi N$ partial wave amplitudes $S_{11}$, $S_{31}$,
$P_{11}$, $P_{31}$, $P_{13}$, and $P_{33}$ obtained in
Ref.~\cite{Workman:2012hx}, which we refer to as ``WI~08'', up to $w =
1.35 \gev$.

\section{Numerical results}

\begin{figure}[t]
  \centering
  \begin{minipage}{0.55\hsize}
    \Psfig{7.2cm}{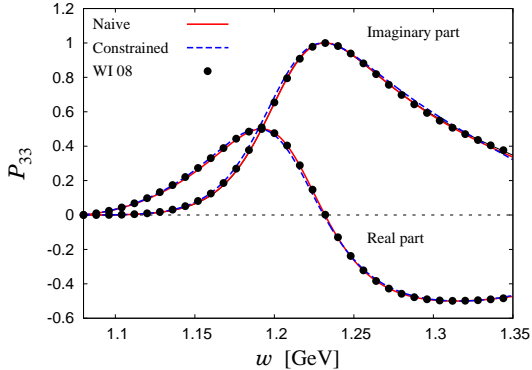}
  \end{minipage}
  \begin{minipage}{0.35\hsize}
    \caption{(color online) Scattering amplitude $P_{33}$ with naive
      (red solid lines) and constrained (blue dashed lines) parameter
      sets fitted to the WI~08 solution (filled
      circles)~\cite{Workman:2012hx}.}
    \label{fig:1}
  \end{minipage}
\end{figure}

\begin{table}[b]
  \caption{Properties of $\Delta (1232)$ and $N(940)$.}
  \label{tab:1}
  \begin{tabular}{llcclcc}
    \hline
    & & \multicolumn{2}{c}{$\Delta (1232)$} &
    & \multicolumn{2}{c}{$N (940)$} \\
    & & Naive & Constrained & 
    & Naive & Constrained \\
    \hline
    $w_{\rm pole}$ [MeV] & & $1209.8 - 47.6 i$ & $1206.9 - 49.6 i$ &
    & $938.9$ & $938.9$ \\
    $g$ [MeV$^{-1/2}$] & & $0.383 - 0.053 i$ & $0.395 - 0.061 i$ &
    & $0.560$ & $0.516$ \\
    $X_{\pi N}$ & & $0.69 + 0.39 i$ & $0.87 + 0.35 i$ &
    & $-0.18\phantom{-}$ & $0.00$ \\
    \hline
  \end{tabular}
\end{table}

Now let us calculate the $\pi N$ compositeness of $\Delta (1232)$ in
the chiral unitary approach.  We fit the model parameters to the $\pi
N$ scattering amplitude WI~08, and we can reproduce the $\pi N$
amplitude very well with $\chi ^{2} / N_{\rm d.o.f.} = 486.3 / 809$.
The best fit for the $P_{33}$ amplitude is shown in Fig.~\ref{fig:1}
as red solid lines (Naive).  From the $\pi N$ amplitude, we can
extract the $\pi N$ compositeness with the formula~\eqref{eq:X}.  The
result of the $\pi N$ compositeness as well as the pole position and
coupling constant is shown in the second and fourth columns in
Table~\ref{tab:1}.  As one can see, the $\pi N$ compositeness for
$\Delta (1232)$ has large real part comparable to unity.  Therefore,
our refined model reconfirms the result in the previous
study~\cite{Aceti:2014ala}, and the result implies large contribution
of the $\pi N$ cloud to the internal structure of $\Delta (1232)$.
However, for $N (940)$, the $\pi N$ compositeness is real but negative
and hence unphysical, because one cannot interpret it as a probability
even for a stable state.  This is because $d G_{L = 1} / d w ( w =
M_{N})$ is positive, which should be negative as the derivative of the
integrand in Eq.~\eqref{eq:G} becomes negative.

In order to resolve this, in addition to $G_{L} ( w = M_{N} ) = 0$ we
constrain the loop function as $d G_{L = 1} / d w (w = M_{N}) \le 0$,
and the fitted amplitude becomes the blue dashed lines (Constrained)
in Fig.~\ref{fig:1} with $\chi ^{2} / N_{\rm d.o.f} = 1239.9 / 809$.
The properties of $\Delta (1232)$ and $N (940)$ are shown in the third
and fifth columns of in Table~\ref{tab:1}.  The properties of $\Delta
(1232)$ shift only slightly and the $\pi N$ compositeness for $N
(940)$ is non-negative.  Again we reconfirm the result for $\Delta
(1232)$ in the previous study~\cite{Aceti:2014ala}.

Finally we note that there is ambiguity in calculating the $\pi N$
compositeness $X_{\pi N}$ with the loop function in the dispersion
relation~\eqref{eq:G}.  Namely, as discussed in
Ref.~\cite{Hyodo:2008xr}, we can consider a shift of the subtraction
constant $\tilde{A}$, which can be compensated by the corresponding
shift of the interaction $V$ so as not to change the full amplitude
$T$.  This shift of the subtraction constant can change the value of
$d G_{L = 1} / d w$ and hence that of $X_{\pi N}$, since the
subtraction constant survives when we differentiate $G_{L = 1} ( w ) =
( s - M_{N} ) \tilde{A} + (\text{finite part})$. However, if we have a
constraint $d G_{L =1} / d w ( w = M_{N} ) \le 0$, such a shift of the
subtraction constant is also constrained and $d G_{L = 1} / d w$
cannot be close to zero around the $\Delta (1232)$ energy region.  In
particular, in the present calculation $\tilde{A}$ takes the maximal
value under the constraint $d G_{L =1} / d w ( w = M_{N} ) \le 0$, as
seen from $X_{\pi N} = 0$ for $N (940)$, which means $d G_{L =1} / d w
( w = M_{N} ) = 0$.  As a consequence, the present calculation would
give a minimal value of $| X_{\pi N} |$ for $\Delta (1232)$ in our
approach from the viewpoint of the shift of the subtraction constant.

\section{Summary}

In this study we have investigated the internal structure of $\Delta
(1232)$ in terms of the $\pi N$ compositeness, which was extracted
from the elastic $\pi N$ scattering amplitude in the chiral unitary
approach.  Fitting the model parameters so as to reproduce the
solution of the $\pi N$ partial wave analysis, we have obtained the
large real part of the $\pi N$ compositeness comparable to unity for
$\Delta (1232)$ and non-negligible imaginary part as well.  Therefore
our refined model reconfirms the result in the previous study on the
$\pi N$ compositeness for $\Delta (1232)$.  This implies large
contribution of the $\pi N$ cloud to the internal structure of $\Delta
(1232)$.
The details of the present study
will be given in a forthcoming paper~\cite{Arai:2015}.

The authors acknowledge A.~Hosaka, H.~Nagahiro, T.~Hyodo, and D.~Jido
for fruitful discussions.  This work is partly supported by the
Grants-in-Aid for Scientific Research from MEXT and JSPS
(No.~15K17649, 
No.~15J06538
).

\end{document}